\documentclass[prb,showpacs,twocolumn,preprintnumbers,amsmath,amssymb,floatfix]{revtex4}
\usepackage[dvips]{graphicx}
\usepackage{color}
\usepackage[latin1]{inputenc}
\usepackage{stmaryrd}
\usepackage{wasysym}
\usepackage{subfigure}

\begin{document}
\draft
\title{Evidence of unconventional low-frequency dynamics in the normal phase of Ba(Fe$_{1-x}$Rh$_x$)$_2$As$_2$ iron-based supercondutors}
\author{L. Bossoni,$^{1,2}$ P. Carretta,$^{1}$  W. P. Halperin,$^{3}$ S. Oh,$^{3}$ A. Reyes,$^{4}$ P. Kuhns,$^{4}$ P.C.
Canfield$^{5}$.}
\address{$^{1}$ Department of Physics, University of
Pavia-CNISM, I-27100 Pavia, Italy}
\address{$^{2}$ Department of Physics ``E. Amaldi,'' University of Roma Tre-CNISM, I-00146 Roma, Italy}
\address{$^{3}$ Department of Physics and Astronomy, NorthWestern University, Evanston, IL, USA}
\address{$^{4}$ National High Magnetic Field Laboratory, Tallahasse, FL, USA}
\address{$^{5}$ Ames Laboratory US DOE and Department of Physics and
Astronomy, Iowa State University, Ames, IA 50011, USA}

\begin{abstract}
This work presents $^{75}$As NMR spin echo decay rate ($1/T_2$)
measurements in Ba(Fe$_{1-x}$Rh$_x$)$_2$As$_2$ superconductors,
for $0.041\leq x\leq 0.094$. It is shown that $1/T_2$ increases
upon cooling, in the normal phase, suggesting the onset of
an unconventional very low-frequency activated dynamic. The
correlation times of the fluctuations and their energy barriers
are derived. The motion is favored at large Rh content, while it
is hindered by the application of a magnetic field perpendicular
to the FeAs layers. The same dynamic is observed in the
spin-lattice relaxation rate, in a quantitatively consistent
manner. These results are discussed in the light of nematic
fluctuations involving domain wall motion. The analogies with
the behaviour observed in the cuprates are also outlined.

\end{abstract}

\pacs{74.70.Xa, 75.25.nj, 74.20.Mn}
\maketitle

\draft

%%%%%%%%%%%%%%%%%%%
\narrowtext
%%%%%%%%%%%%%%%%%%%%

%%%%%%%%%%%%%%INTRO%%%%%%%%%%%%%%%%%%%%%%%
The study of the excitations in the normal phase of
superconductors (SC) is of major importance to unravel the mechanisms
driving the Cooper pair formation. Both in the cuprates and in the
iron pnictides the presence of competing interactions gives rise
to complex phase diagrams and to quasi-degenerate ground-states,
which can induce unconventional dynamics at low energies.
Nuclear Magnetic Resonance (NMR) has played a key role in the
study of the low-frequency (LF) excitations in the normal state both of
high T$_c$ SC and, more recently, of the iron-based
SC. Most of the NMR investigations carried out so far
in these materials have concentrated on the dynamical features
emerging from the spin-lattice relaxation rate ($1/T_1$)
measurements,\cite{Kitagawa2008,Kitagawa2010,Fu2012,Ning2010,Torchetti,Parker2008,Oh2012}
while less attention has been paid to the study of the spin-echo
decay rate $1/T_2$, which is quite a useful tool to probe very
LF excitations.\cite{Bossoni2012,Oh2011,Xiao2012} In
the cuprates, one of the most significant achievements was the
derivation of the staggered static spin susceptibility from
$^{63}$Cu(2) Gaussian echo decay rate
$1/T_{2G}$.\cite{Takigawa1994,PS1991} However other nuclei show a different behavior of
$1/T_2$: $^{89}$Y NMR decay rate, in the SC phase of YBa$_2$Cu$_3$O$_7$, presents an
exponential term,\cite{Suh1993,Recchia1997} which reveals a peak in $1/T_2$, that was firstly ascribed to vortex dynamics.\cite{Suh1993}
Remarkably, $^{17}$O NMR revealed a second peak that, however, was also seen in $^{63}$Cu NQR experiments in YBa$_2$Cu$_3$O$_{7-x}$, where no
magnetic field was applied, thus questioning the former explanation and suggesting other mechanisms involving charge
fluctuations.\cite{Kramer1999,Recchia1997,Bondar1989}

Similar trends of $1/T_2$ have been reported also in  the recently
discovered iron-based SC. In the optimally doped
Ba(Fe$_{1-x}$Co$_x$)$_2$As$_2$\cite{Oh2011} and in
Ba(Fe$_{1-x}$Rh$_x$)$_2$As$_2$ (BaFeRh122
hereafter),\cite{Bossoni2012} a peak in $1/T_2$ was detected below
$T_c$ and again associated with the vortex dynamics. On the other
hand, the behavior of $^{75}$As NMR $T_2$ in the normal phase of
BaFeRh122 SC is not completely understood
\cite{Oh2011,Bossoni2012} and its magnitude is far from any
theoretical expectation.

In the following, a systematic study of $^{75}$As NMR spin echo
decay in BaFeRh122 iron-based SC, over a
broad range of Rh doping, is presented. The echo shows a high temperature (T)
dominant Gaussian decay which becomes exponential at low T.
%Such double
%component was also remarkably found in $^{199}$Hg NMR study of
%HgBa$_2$CuO$_{4+\delta}$,\cite{SuhPHD} as well as $^{63}$Cu NMR
%study of YBa$_2$Cu$_3$O$_y$.\cite{Wu2011}
The exponential decay rate increases upon cooling already in the
normal phase, suggesting the onset of an unconventional very
LF activated dynamics, whose characteristic correlation
times are derived together with the
corresponding energy barriers. This dynamic persists across
the whole phase diagram up to the overdoped compounds, but it is
less pronounced if the magnetic field is applied perpendicularly
to the FeAs layers. It is also shown that the same dynamics affect
$1/T_1$. These LF fluctuations are discussed in terms
of domain wall motion, possibly involving nematic fluctuations.\\

\begin{figure}[htbp]
\includegraphics[width=8cm, keepaspectratio]{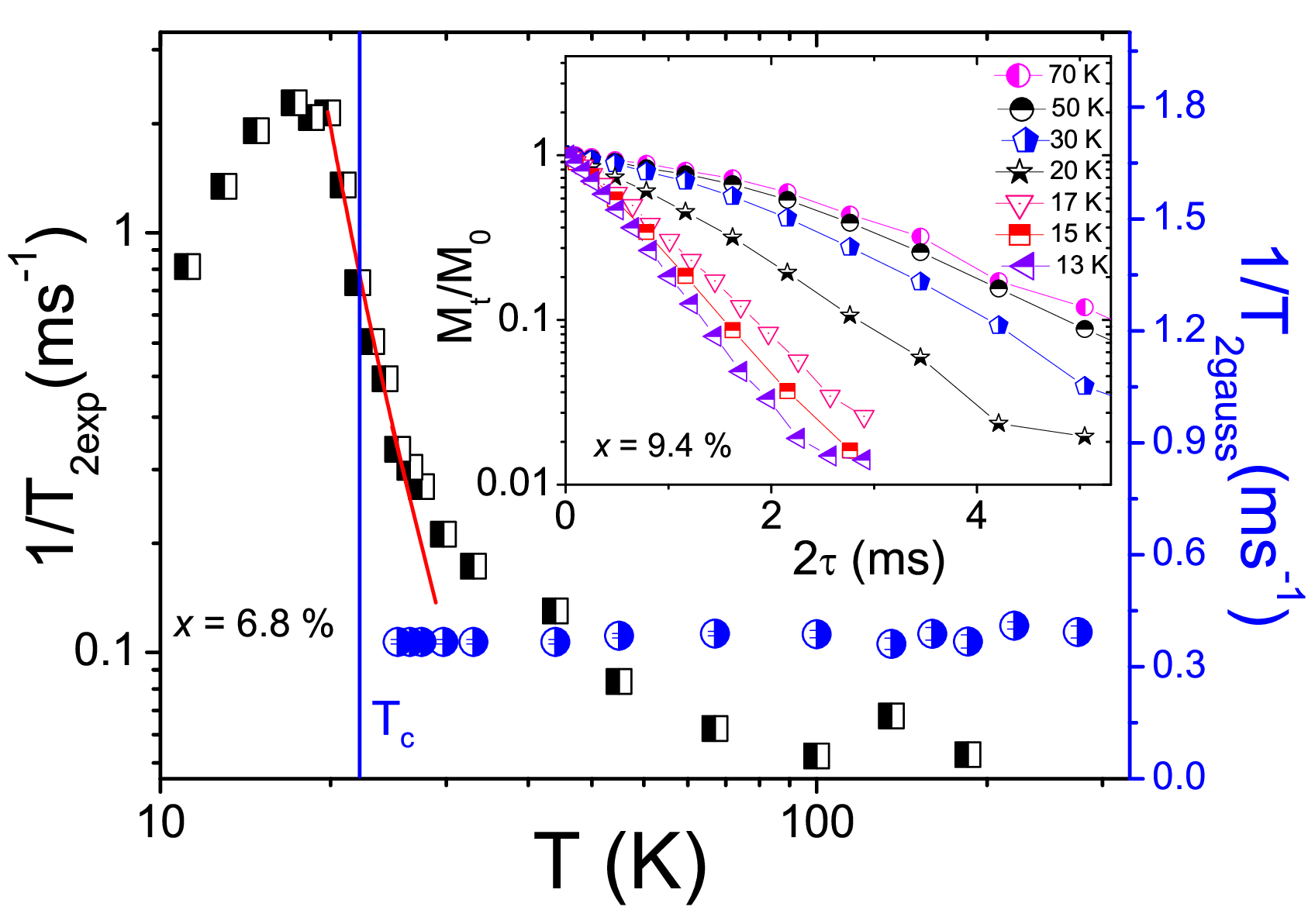}
\caption{The exponential (black half filled squares) and Gaussian
(blue half filled circles) $1/T_2$ measured by Hahn echo at 11 T
$\parallel c$ of the $x=$6.8\% sample. The red solid line shows
the best fit to fast motion equation (see text). (Inset) The spin-echo
amplitude decay at different T, for $x =$ 9.4\%, at
$\mathbf{H_0}=11$ T $\parallel c$ axis, corrected by the $T_1$
contribution. } \label{Fig1}
\end{figure}

\begin{figure}[!h]
\includegraphics[height=8cm, keepaspectratio]{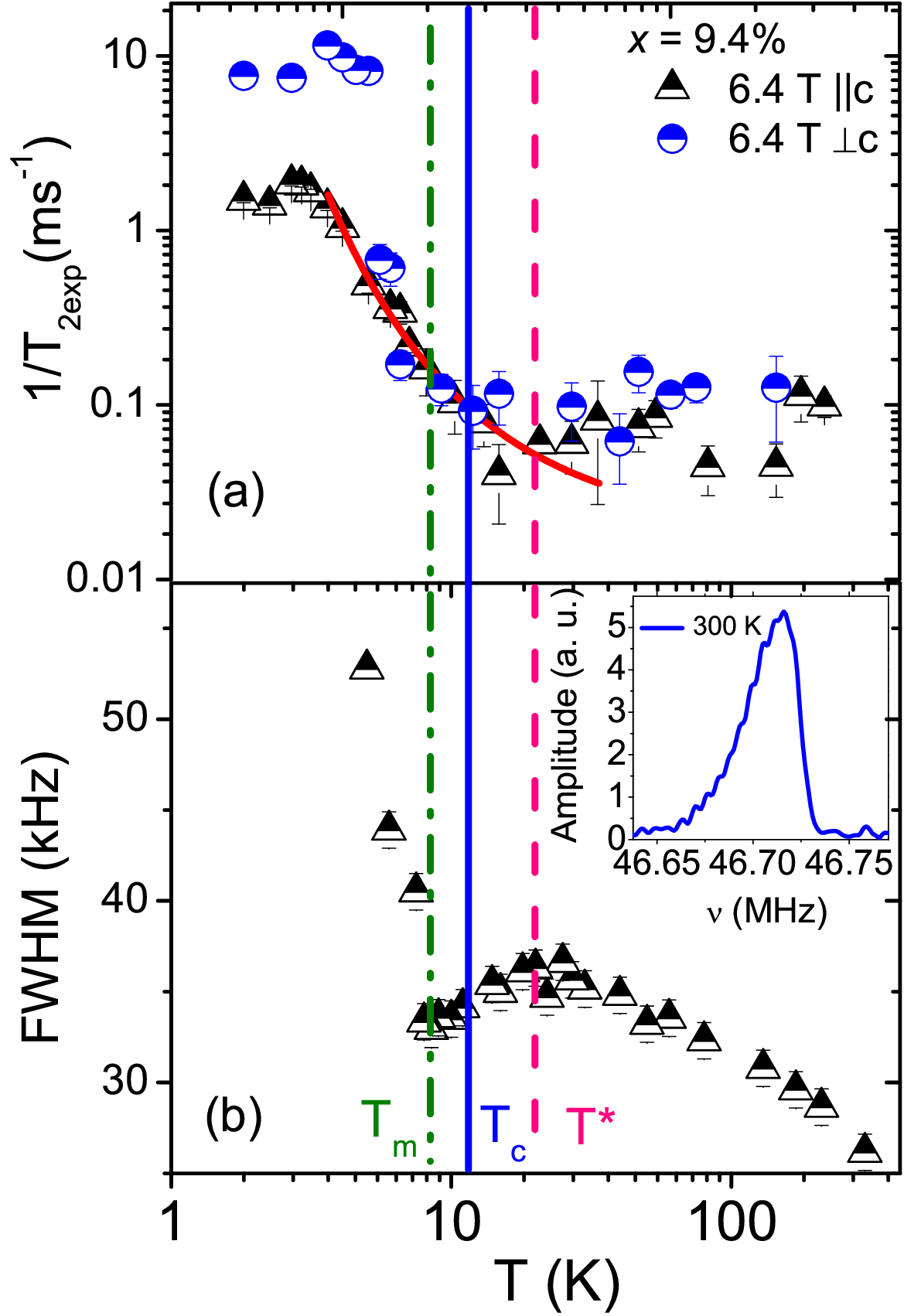}
\caption{\textbf{(a)} $1/T_{2exp}$ measured at $H_0=$ 6.4 T
$\parallel c$, for $x=$9.4\%. The red solid line shows the best
fit according to the fast motion equation (see text). The blue circles show $1/T_{2exp}$
measured at $H_0=$ 6.4 T $\perp c$. \textbf{(b)} The T dependence
of the FWHM of $^{75}$As  central line: the width increases with
decreasing T down to $T^*$ (red dashed line), where it starts to
decrease. Finally the linewidth increases again at the vortex
freezing temperature $T_m$ (green dashed-dotted line). The blue
line marks $T_c$ at 6.4 T. The inset displays an example of the
$^{75}$As NMR spectrum, showing a clear asymmetry. } \label{Fig2}
\end{figure}

\begin{figure}[h!]
%\subfigure%
%{
\includegraphics[width=7.2cm, keepaspectratio]{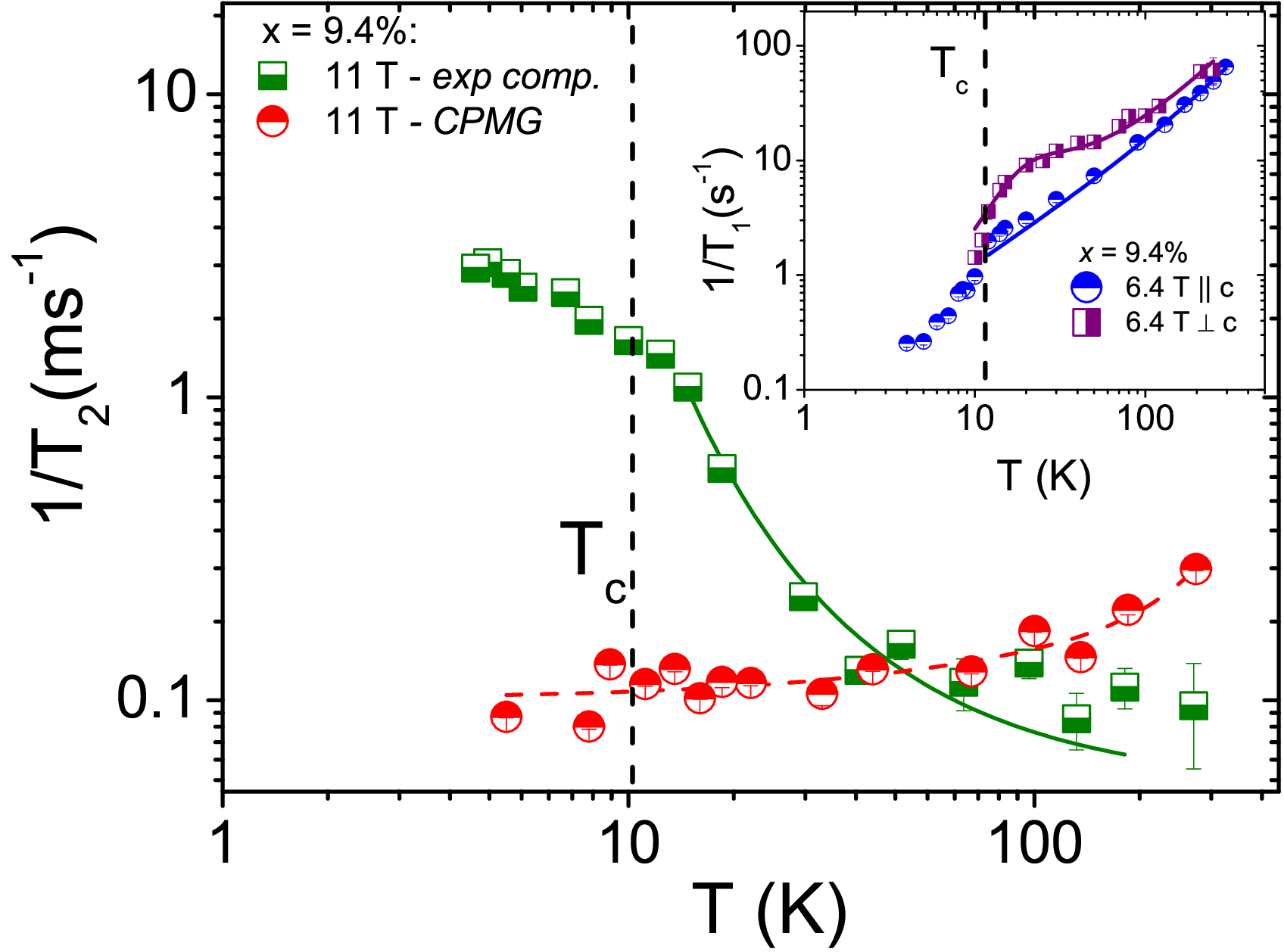}%}%\qquad\qquad
%\subfigure%
%{
%\includegraphics[height=7cm, width=7cm, keepaspectratio]{Fig2b.eps}}%\qquad\qquad
\caption{The \textit{HE} $1/T_{2exp}$ is shown for the overdoped
sample, at 11 T $\parallel c$ (green half filled squares). The red
half filled circles show the $1/T_{2CPMG}$ at 11 T $\parallel c$.
The green solid line is the best fit according fast motion equation,
while the dashed line is a guide-to-the-eye. The dashed vertical
line indicates $T_c$. The inset displays $1/T_1$ data collected at
6.4 T $\parallel$ (blue circles) and $\perp$ (purple squares) to
the $c$ axis. The solid lines are $1/T_1$ best fits according to a
power law (blue line) or to the sum of a BPP and a power law
(purple line). } \label{Fig3}
\end{figure}

NMR measurements have been performed on three BaFeRh122 single
crystals: an underdoped sample, with x = 4.1\% ($T_c=13.6$
K), a nearly optimally
doped sample, with x = 6.8\% ($T_c=22.4$ K), and an overdoped
sample, with x = 9.4\% ($T_c=15.1$ K). The samples were grown as outlined in Ref. \onlinecite{Ni2008}. The critical temperature
$T_c$ was determined via Superconductor Quantum Interference
Device (SQUID) magnetometry and it is in agreement with Ref. \onlinecite{Ni2008}. $^{75}$As NMR experiments were
performed at 6.4 T, 9 T and 11 T, for $\mathbf{H_0}$ parallel
and perpendicular to the $c$ axis. The spin echo decay time was
estimated by fitting the decay of the transverse nuclear
magnetization $M_t$, measured either after a standard Hahn echo
(\textit{HE}) sequence $(\pi/2-\tau-\pi)$ or by using a
Carr-Purcell-Meiboom-Gill (\textit{CPMG}) sequence
$(\pi/2_x-\tau_{CP}-\pi_y-\tau_{CP}-\pi_y...)$. In the latter, the
delay $\tau_{CP}$ was varied in order to extract the intrinsic
decay time $T_{2CPMG}$ for $\tau_{CP}\rightarrow 0$. The
\textit{HE} decay was first corrected in order to remove the
contribution of the spin-lattice relaxation rate
$1/T_1$.\cite{WC1995} Afterwards, the echo decay could
be nicely fit by the product of an exponential and a Gaussian
decay (inset of Fig. \ref{Fig1}):
\begin{equation}
M_t(2 \tau)/M_0=\exp(-2 \tau/T_{2exp})\exp(-(2 \tau)^2/2T_{2G}^2).
\end{equation}
The two components evolve with temperature in such a way that, in
the high T regime, the Gaussian term is significantly larger than
the exponential one, and both are weakly temperature dependent.
This trend persists down to a temperature $T^*>T_c$, where the
Gaussian contribution becomes negligible, while the exponential
rate grows rapidly and becomes the main contribution to the echo
decay (Figs. \ref{Fig1} and \ref{Fig2}). The experimental values
of $T^*$ are 22 $\pm 2$ K and $18\pm 2$ K for $x=6.8$\% and
$x=9.4$\%, respectively. Here the low T
exponential component is discussed, since the high T constant behavior has been
discussed by Oh et al.\cite{Oh2011} From Fig.\ref{Fig3} one
notices that also $1/T_{2CPMG}$ is significantly reduced with
respect to $1/T_{2exp}$ for $T<T^*$ and shows a less pronounced
field and T dependence. The \textit{HE} and \textit{CPMG}
sequences are most sensitive to fluctuations with a characteristic
time scale $\tau_c\sim \tau$ and $\sim \tau_{CP}$, respectively.
Hence, the observed difference between $1/T_{2CPMG}$ and
$1/T_{exp}$ suggests that the correlation time of the fluctuations probed by $^{75}$As
nuclei increases above the $\mu s$, below $T^*$. By comparing the
measurements performed on the three crystals, for different
magnetic field orientations (Fig. \ref{Fig2}) and magnitudes
(Figs. \ref{Fig2} and \ref{Fig3}), one can conclude that the
enhancement of $1/T_{2exp}$ has four main features: (i) it starts
above $T_c$, (ii) it is favored by the magnetic field, (iii) it is
accentuated for in-plane fields and (iv) it persists across the
whole phase diagram, up to the overdoped compound (Fig.
\ref{Fig3}).

%%%%%%%%%%%%%%%%%%%%%%%%%%%%%%
%%%%%%%%%%%%%%%%%%%%%%%%%%%%%%%%%

Further insights into the LF dynamics can be
derived from the T-dependence of the full width at half maximum
(FWHM) of the central $1/2\rightarrow -1/2$ NMR line, obtained
from the Fourier transform of half of the echo (Fig. 2). The NMR
spectrum displays an inhomogeneously broad lineshape with a
LF tail, and it gets broader upon cooling. Remarkably,
at the same temperature $T^*$  where the echo decay becomes
exponential, the linewidth starts to decrease, hence suggesting
the onset of LF dynamics which can average-out the
static frequency distribution probed by $^{75}$As nuclei. Finally,
at $T_m< T_c$, the line broadens again (Fig. \ref{Fig2}), as
expected when the solid/glassy vortex phase sets
in.\cite{Bossoni2012,Bossoni2013,Oh2011}

%%%%%%%%%%%%%%%%%%%%%%%%%%%%%%
%%%%%%%%%%%%%%%%%%%%%%%%%%%%%%%%%

\begin{figure}[h!]
\includegraphics[width=6cm, keepaspectratio]{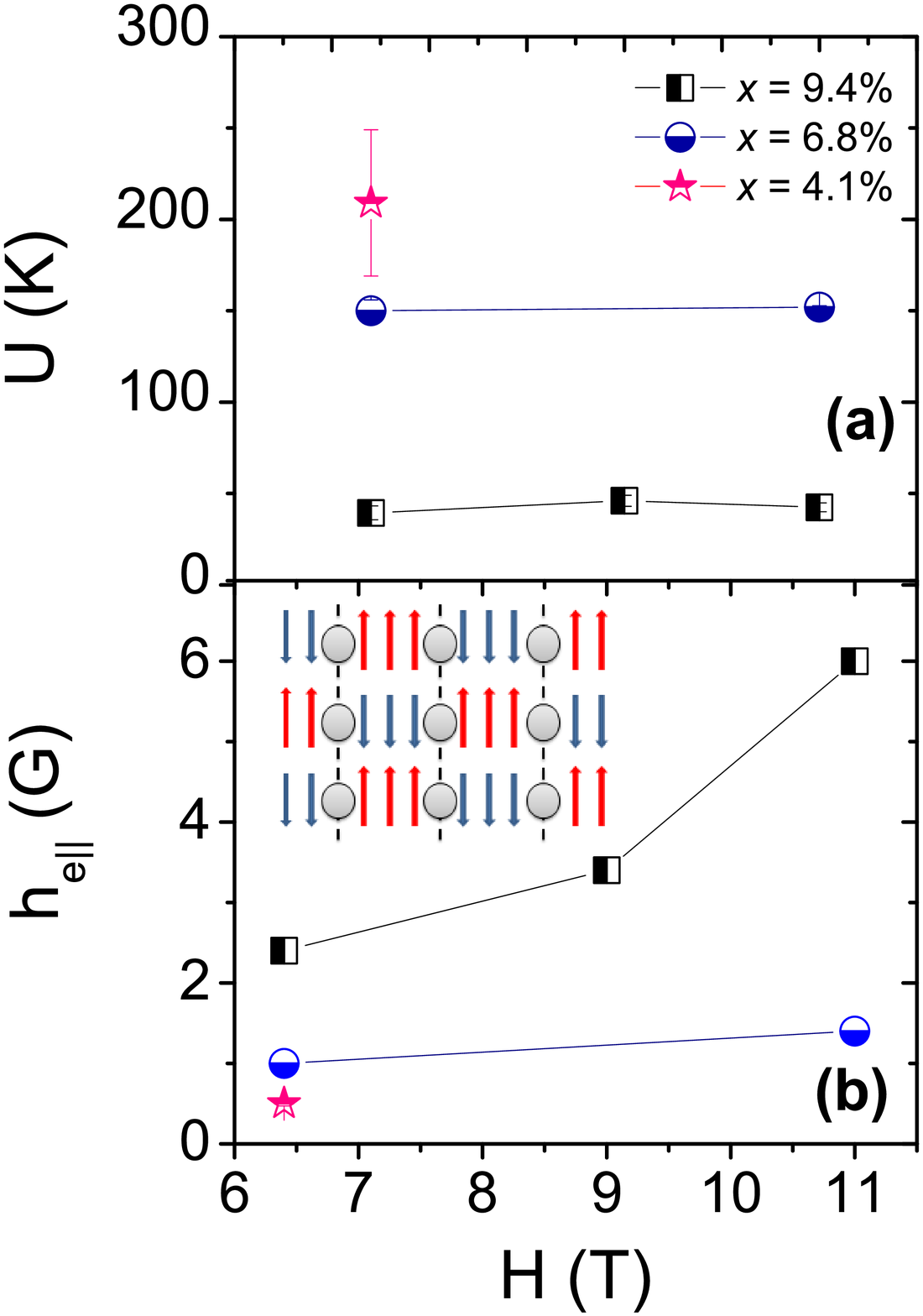}
\caption{\textbf{(a)} Energy barriers
estimated from $1/T_{2exp}$ according to fast motion equation at different
doping levels. \textbf{(b)} The field dependence of the amplitude
of the longitudinal field fluctuations at different Rh
contents. The inset shows a sketch where columnar
antiferromagnetic regions are separated by \textit{antiphase}
domain wall: the blue/red arrows stand respectively for the
down/up spins, while the grey circles indicate the electronic
charges which may favor the domain formation.} \label{Fig4}
\end{figure}

Since a LF dynamic is present one should expect an
effect also on $1/T_1$, which probes the spectral density at the
nuclear Larmor frequency. In the inset of Fig. 3 it is shown
$1/T_1$, as derived after a saturation recovery pulse sequence,
for both magnetic field orientations, for $x= 9.4$\%. Remarkably,
a bump in the spin-lattice relaxation rate was observed in the
normal state, when $\mathbf{H_0 \perp}$ c, which is nearly absent
for $\mathbf{H_0 \parallel}$ c. Notice that the corresponding
$1/T_1T$ data for $\mathbf{H_0 \perp}$ c are quantitatively in
agreement to those measured by Ning et \textit{al}.
\cite{Ning2010} in Ba(Fe$_{1-x}$Co$_x$)$_2$As$_2$ crystals for the
same magnetic field orientation.
The peak behavior of $1/T_1$ for the two orientations is suggestive of the Bloembergen-Purcell-Pound (\textit{BPP}) mechanism,\cite{Abragam} accounting for a LF
activated dynamic (Fig. \ref{Fig4}), plus a power law trend
$T^{\alpha}$ ($\alpha\rightarrow 1$), which characterizes these
compounds when $\mathbf{H_0} \parallel$ c. A fit to this model results into an energy barrier $U=50$ $\pm$ 5 K corresponding to a correlation time at infinite temperature $\tau_0=5.2$ x $10^{-10}s$ and an average fluctuating field $h_{e\perp}=19.4\pm 2.2$ G.
Such parameters are comparable with
the ones recently reported by Hammerath et \textit{al.} in
underdoped LaO$_{1-x}$F$_x$FeAs.\cite{Franzi2013} In principle,
two possible reasons for the anisotropy in $1/T_1$ should be
considered: (i) the role of the magnetic field in inhibiting the
LF fluctuations when $\mathbf{H_0}\parallel$ c, (ii)
the filter effect of the hyperfine form factor.\cite{Shannon2011}
Given the behaviour found for $1/T_{2exp}$ for the different
magnetic field orientations (see Fig. 2a) the former scenario
appears more likely.
%%%%1/T2%%%

%%%%%%%%%%%%%%%%%%%%%%%%%%%%%%%%%

Further quantitative information on the correlation times
describing the LF activated dynamics can be gained from
the analysis of the spin echo decay rate $1/T_{2exp}$. When the
\textit{HE} becomes exponential, the spin-echo relaxation rate can be fit by the
fast motions expression\cite{Takigawa}
\begin{equation}
1/T_{2exp}= ^{75}\gamma^2<h_{e\parallel}^2>\tau_0
e^{U(H,x)/T},
\end{equation} 
where the energy barrier $U$ is assumed to depend on the field
intensity $H_0$ and on the Rh concentration $x$. The fit results,
obtained by using the $\tau_0$ value derived from $1/T_1$, are
shown in Fig. \ref{Fig4}.
 %%%%%%%%%%%%%%%%%%%%%%%%%%%%%%
%%%%%%%%%%%%%%%%%%%%%%%%%%%%%%%%%
The reader may notice that the barrier is comparable with the one
found from the $1/T_1$ fit. Moreover $U$ is $H_0$-independent, in
the explored range, while it clearly depends on the electronic
concentration, namely it decreases by increasing the Rh content.
This trend indicates that the higher the electron doping the
faster the dynamics. Furthermore the fluctuating longitudinal
local field $h_{e\parallel}$ shows a continuous increase with the applied field and the doping.
It is also noticed that, when the field is perpendicular to the $c$ axis, the
enhancement is significantly larger.
%Accordingly,
%the observation of the bump in $1/T_1$ solely for
%$\mathbf{H_0}\perp$ c suggests that the magnetic field is playing
%an active role in reducing the spin dynamics for
%$\mathbf{H_0}\parallel$ c.

%\begin{equation}
%\frac{1}{T_{2CPMG}}=\frac{2\tau}{T_{2g}}+\frac{1}{T_{2exp}}
%\end{equation}

The LF dynamics evidenced by $1/T_{2exp}$ and by the
bump in $1/T_1$ cannot be due to the standard correlated electron
spin fluctuations or SC fluctuations which
typically occur at frequencies orders of magnitude larger than the frequency probed here. One should look for very
LF fluctuations as the ones occurring close to a spin
or charge freezing or taking place among quasi-degenerate
ground-states. In this respect, one should consider that, owing to
the geometry of the relevant exchange couplings, the magnetic
properties of the iron pnictides have often been described within
an effective J$_1$-J$_2$ model on a square lattice.\cite{Si2008}
The ground-state of that model is characterized by two degenerate
columnar antiferromagnetic ground-states corresponding to  two
nematic phases, and the fluctuations between those two states can
give rise to very LF dynamics. In fact in vanadates,
which can be considered as prototypes of that model, these
LF fluctuations have been detected by $\mu$SR above the
magnetic ordering,\cite{Carretta2002} in a temperature range where
the electron spins are already correlated.\cite{Bossoni2011} These
dynamics can be associated with domain wall motion separating
correlated regions of the two nematic states. Once the two
different phases set in, the domain walls can be put into motion
if the energy barrier $U$ separating these two
phases\cite{Chandra1990} is overcome (Fig. \ref{Fig4}). Notice
that a similar scenario has also been proposed in a recent study
of the magnetic state of
CaFe$_2$As$_2$.\cite{Xiao2012}\\
On the other hand, %\textcolor{red}{it is important to notice that the behaviour
%observed here shows similarities with the behaviour found in the
%cuprates, where nematic fluctuations are not expected. In fact, an
%analogous behavior has also been detected in Hg-based high T$_c$
%superconductors.}
the observations of the magnetic field effect on $1/T_2$ and
$1/T_1$, suggest that when $\mathbf{H_0}\parallel$ c the fluctuations
are reduced. Such an effect recalls the field-induced charge order
recently reported in underdoped YBa$_2$Cu$_3$O$_{6+x}$
SC, where the blocking of the charge order occurs
only if $\mathbf{H_0}$ is perpendicular to the highly conductive
CuO$_2$ layers.\cite{Wu2011,Wu2013} Moreover in these systems, Wu
et \textit{al.}\cite{Wu2011} found a temperature and magnetic
field dependence of $1/T_2$ similar to the one reported here
for the iron-based SC. In the light of these
analogies one could speculate that very LF fluctuations
associated with domain wall motion, possibly involving charge
stripes, can be present both in the cuprates and in the iron-based
SC. However, further experiments are required to
support such a scenario.

In conclusion, this paper presents a systematic study of the
spin-echo decay rate in 122 iron-pnictides over a broad doping
range. A LF spin dynamic is observed above $T_c$ and it
is responsible for a bump in the spin-lattice relaxation time, as
well as for an enhancement of the exponential component of the
spin-echo decay rate. Such dynamics are common at all the doping
concentrations, and they get faster in the overdoped regime. Moreover they can be associated with domain wall motion, possibly
involving nematic fluctuations. The remarkable analogies with the
behavior found in the cuprates indicate the need for a deeper
investigation, also by other techniques.\\
We would like to thank A. Mounce, J. P. Lee, F. Hammerath and S.-J. Yuan for help and useful discussion. The research activity in Pavia was supported by Fondazione Cariplo
(Research Grant No. 2011-0266). Research at N. U. was supported by the U.S. DOE, Office
of Basic Energy Sciences (BES), Division of Materials Sciences and
Engineering (MSE), award DE-FG02-05ER46248. Work done in Ames Laboratory (P.C.C.) was supported by the U.S. DOE, BES Office, MSE Division under contract No. DE-AC02-07CH11358. A portion of this work was performed at the NHMFL, which is supported by National Science Foundation Cooperative Agreement No. DMR-0654118, the State of Florida, and the U.S. DOE.

%%%%%%%%%%%%%%%%%%%%%%%%%%%%%

\end{document}